# Mexican hat curve for hydrogen and antihydrogen-states in natural atom H.[1]

G. Van Hooydonk, Ghent University, Faculty of Sciences, Krijgslaan 281, B-9000 Ghent (Belgium)
(guido.vanhooydonk@rug.ac.be)

**Abstract**

*Molecular band spectra as well as atomic line spectra reveal a left-right symmetry for atoms (Van Hooydonk, Spectrochim. Acta A, 2000, **56**, 2273 and submitted, 2002). We now extract a Mexican hat shaped or double well curve from the line spectrum (Lyman ns½ singlets) of natural atom H. An H CSB theory and its oscillator contribution $(1-½\pi/n)^2/n^2$ lead to unprecedented results for antihydrogen physics, ahead of the CERN AD-project on artificial antihydrogen.*

Molecular band [1] and atomic line spectra [2] point towards intra-atomic left-right symmetry or to *atom-antiatom symmetry* [1,2]. This makes sense only if chiral symmetry breaking (CSB) is conform to a Mexican hat shaped energy curve. Double well potentials show 2 minima for stable left and right (or atom and antiatom) states respectively, separated by a maximum between the 2 minima. Earlier [2], we identified a harmonic Rydberg in atom H, left unnoticed thus far and equal to

$$R_{H(harm)} = 109{,}679.3522824 \text{ cm}^{-1} \qquad (1)$$

Using Bohr $1/n^2$ theory with Rydberg (1) and adding negative one-electron energies, we get the net chiral energy contributions $C_{nH}$ for level n

$$C_{nH} = 109{,}679.3522824/n^2 + E_{nH} \qquad (2)$$

As in [2], we use Erickson's QED level energies for H Lyman ns½ singlets [3]. Results are in Table 1.
In atomic CSB theory [2], $C_{nH}$ varies as

$$C_{nH} \sim -(1-½\pi/n)^2/n^2 \qquad (3)$$

with a maximum at $n=\pi$ and minima for $n=½\pi$ and for (trivial) $n=\infty$. The maximum at $n=\pi$ divides the ns singlets in 2 subsets: $n<\pi$ and $n>\pi$ respectively, which are classified as the left/right, atom/antiatom states in natural atom H [2].
Conform (3), $C_{nH}$-values are plotted versus symmetry breaking variable $\pi/n-1$ in Fig. 1.

Table 1. $C_{nH}$ for Lyman ns singlets (in cm$^{-1}$)

| n | π/n-1 | $C_{nH}$ |
|---|---|---|
| 1 | 2.14159 | 0.5785784000 |
| 2 | 0.57080 | 0.0202354000 |
| 3 | 0.04720 | 0.0444608444 |
| 4 | -0.21460 | 0.0406722600 |
| 5 | -0.37168 | 0.0332103960 |
| 6 | -0.47640 | 0.0267241111 |
| 7 | -0.55120 | 0.0216769122 |
| 8 | -0.60730 | 0.0178202625 |
| 9 | -0.65093 | 0.0148561305 |
| 10 | -0.68584 | 0.0125484040 |
| 11 | -0.71440 | 0.0107254237 |
| 12 | -0.73820 | 0.0092646378 |
| 13 | -0.75834 | 0.0080783517 |
| 14 | -0.77560 | 0.0071031351 |
| 15 | -0.79056 | 0.0062924648 |
| 16 | -0.80365 | 0.0056117521 |
| 17 | -0.81520 | 0.0050349169 |
| 18 | -0.82547 | 0.0045420351 |
| 19 | -0.83465 | 0.0041176929 |
| 20 | -0.84292 | 0.0037498300 |

The 4$^{th}$ order polynomial fit with $x=(\pi/n-1)$
$$C_{nH}= 0.04481240x^4 + 0.00018359x^3\ 0.089544361x^2$$
$$- 0.00014286x + 0.04466724 \qquad (4)$$

has been extrapolated to the left in Fig. 1. The perfect double well or Mexican hat shaped energy curve is completely in agreement with our CSB explanation for the line spectrum of natural atom H [2]. States with $n<\pi$ have a different minimum than states with $n>\pi$ and are separated by the expected maximum at $n=\pi$. Fig. 1 basically illustrates the atom-antiatom symmetry in nature. The Universe is not at all matter anti-symmetric, as we suggested some time ago [5].

---
[1] To be presented at the International Conference on *Precision Physics of Simple Atomic Systems* (PSAS) 2002, St. Petersburg, 06/30-07/04 (Editors: K. Jungmann, S.G. Karshenboim and V.B. Smirnov) and at the Wigner Centennial Symposium, 2002 (Pecs, Hungary) 07/09-07/12

Harmonic (achiral) Rydberg (1) in Bohr $1/n^2$ theory generates this simple and classical result. It informs us directly about the existence of *non-annihilating* hydrogen and antihydrogen states in natural atom H [1,2]. Probably [2], the unprecedented result in Fig. 1 even runs ahead of the CERN AD-experiment on *artificially produced antihydrogen* [4].

We must identify the origin of a physical difference between atom and antiatom states, as suggested in [2] and in (4) or in Fig. 1. We can link the Mexican hat curve and (4) with the cut-off in the photon frequency, important for bound state QED and for the chiral behavior of particles [6] as we argued elsewhere [7]. Maybe, bound state QED cannot yet be validated as it stands [2,7]. More results are available [2,7] and more work is presented soon.

---

Fig. 1. Mexican hat curve in atom H: $C_{nH}$ versus $\pi/n-1$ for H Lyman ns-singlets
(states with $n<\pi$ : ● , with $n>\pi$ : o)

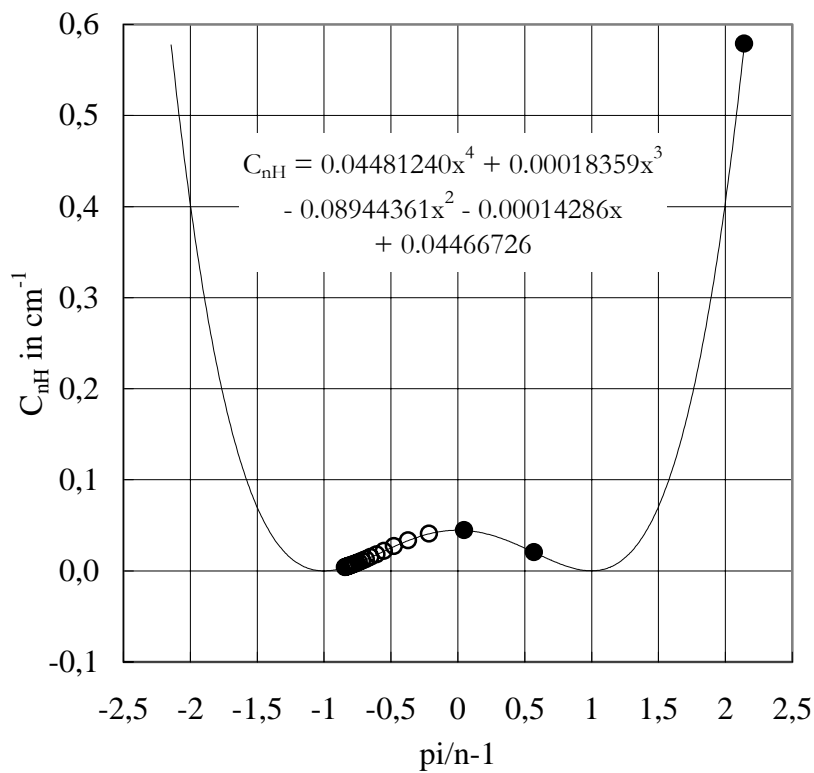

$$C_{nH} = 0.04481240x^4 + 0.00018359x^3 - 0.08944361x^2 - 0.00014286x + 0.04466726$$